\documentclass[
    twoside
]{article}
\usepackage{iwce}
\usepackage{graphicx,calrsfs,amssymb}

\newcommand{\intd}{\textrm{d}}
\newcommand{\intR}{\int_{-\infty}^\infty}

\newcommand{\Eqref}[1]{Eq.~(\ref{#1})}
\newcommand{\Figref}[1]{Fig.~\ref{#1}}

\begin{document}

\title{Modeling of Inelastic Transport in One-dimensional Metallic Atomic Wires}

\IWCEauthorsFirst{THOMAS FREDERIKSEN, MADS BRANDBYGE, ANTTI--PEKKA
JAUHO} \setHeadings{Frederiksen}{Modeling of Inelastic Transport
in One-dimensional Metallic Atomic Wires} \IWCEaddressFirst{MIC --
Department of Micro and Nanotechnology, Technical University of Denmark,\\
{\O}rsteds Plads, Bldg.~345E, DK-2800 Lyngby, Denmark}
\email{thf@mic.dtu.dk}

\IWCEauthorsSecond{NICOL\'{A}S LORENTE}
\IWCEaddressSecond{Laboratorie Collisions, Agr\'{e}gats,
    R\'{e}activit\'{e}, IRSAMC, Universit\'{e} Paul Sabatier,\\
    118 Route de Narbonne, F-31062 Toulouse, France}

\preparetitle

\begin{IWCEabstract}
Inelastic effects in electron transport through nano-sized devices
are addressed with a method based on nonequilibrium Green's
functions (NEGF) and perturbation theory to infinite order in the
electron-vibration coupling. We discuss the numerical
implementation which involves an iterative scheme to solve a set
of coupled non-linear equations for the electronic Green's
functions and the self-energies due to vibrations. To illustrate
our method, we apply it to a one-dimensional single-orbital
tight-binding description of the conducting electrons in atomic
gold wires, and show that this simple model is able to capture
most of the essential physics.
\end{IWCEabstract}

\IWCEkeywords{Inelastic transport, nonequilibrium Green's
functions, self-consistent Born approximation.}

\begin{multicols}{2}
\intro{Introduction} Atomic-size conductors represent the ultimate
limit of miniaturization, and understanding their properties is an
important problem in the fields of nanoelectronics and molecular
electronics. Quantum effects become important which leads to a
physical behavior fundamentally different from macroscopic
devices. One such effect is the inelastic scattering of electrons
against lattice vibrations, an issue which is intimately related
to the important aspects of device heating and stability.

In this paper we describe a method to calculate the inelastic
transport properties of such quantum systems connected between
metallic leads. As a specific example, we here apply it to a
simple model for atomic Au wires, for which such inelastic signals
have recently been revealed experimentally
\cite{AgraitEtAl.2002.OfEDiBAW}.

\section{Inelastic transport formalism}
\noindent Our starting point is a formal partitioning of the
system into a left ($L$) and a right ($R$) lead, and a central
device region ($C$), in such a way that the direct coupling
between the leads is negligible. Hence we write the electronic
Hamiltonian as
\begin{equation}
{\cal H} = {\cal H}_L + {\cal V}_{LC} + {\cal H}_C(q) + {\cal
V}_{RC} + {\cal H}_R,
\end{equation}
where ${\cal H}_\alpha$ is a one-electron description of lead
$\alpha=L/R$ and ${\cal V}_{\alpha C}$ the coupling between
$\alpha$ and $C$. The central part ${\cal H}_C(q)$ is also a
one-electron description but depends explicitly on a displacement
vector $q$ corresponding to mechanical degrees of freedom of the
underlying atomic structure in this region (within the
Born-Oppenheimer approximation we assume instantaneous response of
the electrons). We are here concerned with the electronic
interaction with (quantized) oscillatory motion of the ions. For
small vibrational amplitudes the $q$-dependence can be expanded to
first order along the normal modes $\lambda$ of the structure,
i.e.
\begin{eqnarray}
{\cal H}_{C}({q}) &\approx& {\cal H}_{C}^0 + {\cal
H}_{C}^\mathrm{e-ph},\\
{\cal H}_{C}^0 &=& \sum_{\nu,{\nu'}}H_{\nu,{\nu'}}\hat
c_{\nu}^\dagger
\hat c_{{\nu'}}^{},\\
{\cal H}_{C}^\mathrm{e-ph} &=&
\sum_\lambda\sum_{\nu,{\nu'}}M^\lambda_{\nu,{\nu'}}\hat
c_{\nu}^\dagger \hat c_{{\nu'}}^{}(\hat b_\lambda^\dagger +\hat
b_\lambda^{}),
\end{eqnarray}
where $\hat c_{\nu}^\dagger$ ($\hat c_{\nu}^{}$) is the
single-electron creation (annihilation) operator and $\hat
b_\lambda^\dagger$ ($\hat b_\lambda^{}$) the boson creation
(annihilation) operator. The ionic Hamiltonian is just the
corresponding ensemble of harmonic oscillators
\begin{equation}
{\cal H}_{C}^\mathrm{ion} = \sum_\lambda \Omega_\lambda(\hat
b_\lambda^\dagger\hat b_\lambda^{}+\frac 12),
\end{equation}
where $\Omega_\lambda$ is the energy quantum associated with
$\lambda$.

The transport calculation is based on NEGF techniques
\cite{Haug&Jauho.1996.QKIOS}. For steady state the electrical
current $I_\alpha$ and the power transfer $P_\alpha$ (per spin)
\emph{to} the device from lead $\alpha$ is given by
\cite{TF.2004.thesis}
\begin{eqnarray}
\label{eq:current}
I_\alpha &=& e\langle \dot {\mathcal
N}_\alpha\rangle = \frac
{-e}\hbar \intR \frac{\intd \omega}{2\pi} t_\alpha(\omega),\\
P_\alpha &=& -\langle \dot {\mathcal H}_\alpha\rangle = \frac
1\hbar \intR
\frac{\intd \omega}{2\pi}\omega t_\alpha(\omega), \label{eq:power}\\
t_\alpha(\omega)&\equiv&\textrm{Tr}[\mathbf
\Sigma^{<}_\alpha(\omega)\mathbf G^{>}(\omega) - \mathbf
\Sigma^{>}_\alpha(\omega) \mathbf G^{<}(\omega)],\qquad
\end{eqnarray}
where $\cal N_\alpha$ is the electronic number operator of lead
$\alpha$. Above we have introduced Green's functions in the device
region $\mathbf G^\lessgtr(\omega)$ and the lead self-energies
$\mathbf \Sigma^\lessgtr_\alpha(\omega)$ (scattering in/out rates)
due to lead $\alpha$. For a shorthand notation these are written
as matrices in the $\{\nu\}$-basis. For example, the elements in
$\mathbf G^<(\omega)$ are the Fourier transforms of $G^<(\nu,
t;{\nu'}, t')\equiv i\hbar^{-1}\langle\hat
c^\dagger_{{\nu'}}(t')\hat c^{}_{\nu}(t)\rangle$. In the limit of
zero coupling $M^\lambda_{\nu,{\nu'}}=0$, we can solve exactly for
the lead self-energies $\mathbf
\Sigma_\alpha^{r,\lessgtr}(\omega)$ and the device Green's
functions $\mathbf G_0^{r,\lessgtr}(\omega)$ (since this is a
single-electron problem).

Complications arise with a finite coupling, where the vibrations
mediate an effective electron-electron interaction. To use
\Eqref{eq:current} and (\ref{eq:power}) we need the ``full''
Green's functions $\mathbf G^{r,\lessgtr}(\omega)$. Our approach
is the so-called self-consistent Born Approximation (SCBA), in
which the electronic self-energies due to the phonons $\mathbf
\Sigma_\mathrm{ph}^{r,\lessgtr}(\omega)$ are taken to lowest order
in the couplings \cite{Haug&Jauho.1996.QKIOS}. For a system
lacking translational invariance \cite{TF.2004.thesis}
\begin{eqnarray}
\mathbf \Sigma^{r}_\mathrm{ph}(\omega) &=& i
\sum_{\lambda}\intR\frac{\intd\omega'}{2\pi} \mathbf
M^{\lambda}\Big[\nonumber
\frac{4}{\Omega_\lambda}\mathrm{Tr}[\mathbf
G^<(\omega')\mathbf M^{\lambda}]\\
&&+D^r_0({\lambda},\omega-\omega') [ \mathbf G^<(\omega')+\mathbf
G^r(\omega')]\mathbf
M^{\lambda}\nonumber\\
&&+D^<_0({\lambda},\omega-\omega') \mathbf G^r(\omega')\mathbf
M^{\lambda}\Big], \label{eq:SigmaRet}\\
\mathbf
\Sigma^{\lessgtr}_\mathrm{ph}(\omega) &=&
i \sum_{\lambda}\intR\frac{\intd\omega'}{2\pi} \mathbf M^{\lambda}\nonumber\\
&&\times D^\lessgtr_0({\lambda},\omega-\omega')
    \mathbf G^\lessgtr(\omega')\mathbf M^{\lambda}.
\label{eq:SigmaLessGreat}
\end{eqnarray}
In the above, the phonon Green's functions
$D_0^{r,\lessgtr}(\lambda,\omega)$ are approximated by the
noninteracting ones \cite{Haug&Jauho.1996.QKIOS}. Finally,
$\mathbf G^{r,\lessgtr}(\omega)$ are related to $\mathbf
G_0^{r,\lessgtr}(\omega)$, $\mathbf
\Sigma_\alpha^{r,\lessgtr}(\omega)$, and $\mathbf
\Sigma_\mathrm{ph}^{r,\lessgtr}(\omega)$ via the Dyson and Keldysh
equations \cite{Haug&Jauho.1996.QKIOS}
\begin{eqnarray}
\mathbf G^{r}(\omega) &=& \mathbf G^{r}_0(\omega) +\mathbf G^{r}_0
(\omega) \mathbf \Sigma^{r}_\mathrm{ph}(\omega) \mathbf
G^{r}(\omega),\label{eq:GRet}\\
\mathbf G^{\lessgtr}(\omega) &=& \mathbf G^{r}(\omega) [\mathbf
\Sigma^{\lessgtr}_L+\mathbf\Sigma^{\lessgtr}_R +\mathbf
\Sigma^{\lessgtr}_\mathrm{ph}](\omega)\mathbf
G^{a}(\omega).\label{eq:GLessGreat}\qquad
\end{eqnarray}

The coupled non-linear equations
(\ref{eq:SigmaRet})--(\ref{eq:GLessGreat}) have to be solved
iteratively subject to some constraints on the mode population
$n_\lambda$ (appearing in $D^\lessgtr_0(\lambda,\omega)$). We
identify two regimes: (i) the externally damped limit where the
populations are fixed according to the Bose distribution
$n_\lambda=n_B(\Omega_\lambda)$, and (ii) the externally undamped
limit where the populations vary with bias such that no power is
dissipated in the device, i.e.~$P_L+P_R=0$. To solve the above we
have developed an implementation in \textsc{Python}, in which the
Green's functions and self-energies are sampled on a finite energy
grid.

\section{Simple model}
\noindent As a simple illustration of our method, let us consider
an infinite one-dimensional single-orbital tight-binding chain. We
define the central region $C$ to be a piece of it with $N+2$ sites
to represent the conducting electrons in a finite metallic atomic
wire. The two semi-infinite pieces which surround $C$ can now be
considered as left and right leads. Ignoring on-site energy and
hopping beyond nearest neighbors we simply have for $C$
\begin{equation}
{\cal H}_{C}({q}) = \sum_{i=1}^{N+1}t_{i,i+1}(q)(\hat
c_{i}^\dagger \hat c_{i+1}^{}+\mathrm{h.c.}).
\end{equation}
The hopping amplitudes explicitly depend on the displacement
vector $q$ where the coordinate $q_i$ describes the displacement
of ion $i$ from its equilibrium position. As a specific model for
the hopping modulation by displacement we use the so-called
Su-Schrieffer-Heeger (SSH) model
\cite{Su&Schrieffer&Heeger.1979.SIP} in which the hopping
parameter is expanded to first order in the intersite distance
\begin{equation}
t_{i,i+1}(q) = t^0 +t'({q}_i-{q}_{i+1}),
\end{equation}
where $t^0$ and $t'$ are site-independent parameters. To describe
the ions (in a uniform chain where the end sites are fixed in
space, $q_1=q_{N+2}=0$) we include only nearest neighbor springs
and write
\begin{equation}
{\cal H}^\mathrm{ion}_C = \sum_{i=1}^{N+1} \Big[\frac 12 m\dot
{q}_i^2+\frac 12 K({q}_i-{q}_{i+1})^2\Big],
\end{equation}
where $m$ is the ionic mass and $K$ the effective spring constant
between two neighboring sites.

Imposing quantization via $[q_i,\dot q_j]=i\hbar\delta_{i,j}$, we
can formulate the linearized electron-vibration interaction in
terms of the normal mode operators $\hat b_\lambda$ and $\hat
b_\lambda^\dagger$,
\begin{equation}
{\cal
H}_C^\mathrm{e-ph}=\sum_{\lambda=1}^{N}\sum_{i=1}^{N+1}M_{i,i+1}^\lambda
    (\hat c_{i}^\dagger \hat c_{i+1}^{}+\textrm{h.c.})(\hat b_\lambda^\dagger +\hat
    b_\lambda^{}),
\end{equation}
and relate the coupling elements to components of the normal mode
vectors $\mathbf{e}_\lambda$ (normalized $\mathbf{e}_\lambda\cdot
\mathbf{e}_\lambda=1$) as \cite{TF.2004.thesis}
\begin{eqnarray}
M^\lambda_{i,i+1} &=&t' \hbar ~\frac{(\mathbf{e}_\lambda)_i
-(\mathbf{e}_\lambda)_{i+1}}{\sqrt{2m\Omega_\lambda}}.\qquad
\label{eq:couplings}
\end{eqnarray}

It is well established that atomic Au wires have one almost
perfectly transmitting eigenchannel at the Fermi energy
(e.g.~\cite{AgraitEtAl.2002.OfEDiBAW} and references herein). To
avoid reflection in our model we describe the leads with the same
electronic parameters as for the wire, leading to semi-elliptic
band structures of the leads with widths $4t^0$. With one electron
per site the band is half filled and the Fermi energy becomes
$\varepsilon_F=0$. Further, we take the lead states to be occupied
according to Fermi distributions $n_F(\omega-{\mu}_\alpha)$ where
the chemical potentials vary as ${\mu}_L = +eV/2$ and ${\mu}_R =
-eV/2$. With this information we essentially have $\mathbf
\Sigma_\alpha^{r,\lessgtr}(\omega)$ \cite{TF.2004.thesis}. The
setup and the set of normal modes for a particular $N=6$ atomic
wire are shown in \Figref{fig:setup}.

\begin{IWCEfigure}
  \centering
  \includegraphics*[width=.8\textwidth]{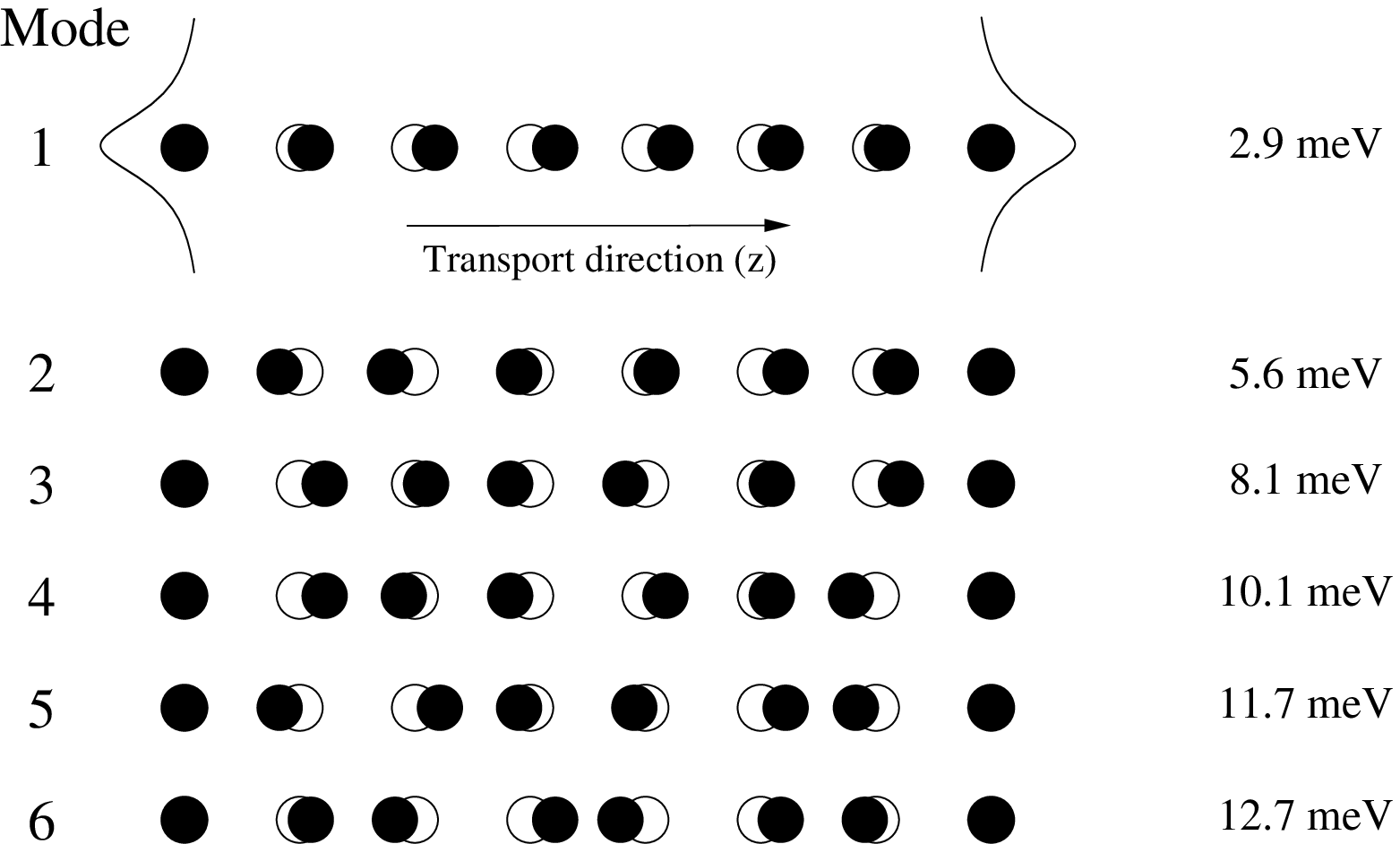}
  \caption{Illustration of the normal modes (longitudinal) of a 6-atom wire
  arranged between two fixed end sites (level-broadened due coupling to
  semi-infinite leads). The open circles represent the
  equilibrium configuration, and the black discs a
  displacement proportional to the normal mode vectors. The
  modes are arranged vertically according to the mode energy
  $\Omega_\lambda$, which are also shown to the right of each
  mode vector ($K=2$ eV/$\mathrm{\AA}^2$). Note that the
  highest energy mode has alternating bond length (ABL) character.}
  \label{fig:setup}
\end{IWCEfigure}

\section{Numerical results}
\noindent Let us now discuss our numerical results for the
differential conductance calculated with \Eqref{eq:current} for
different lengths $N$ and spring constants $K$. We use the
parameter values stated in Tab.~\ref{table1} which qualitatively
yields reasonable agreement with the experimental measurements on
atomic Au wires \cite{AgraitEtAl.2002.OfEDiBAW}.

\begin{IWCEtable}
  \centering
  \begin{tabular}{lcl}
    \hline
    Physical quantity & Symbol & Value\\
    \hline\hline\vspace{-3mm}\\
    Bare hopping & $t^0$ & 1.0 eV\\
    Hopping modulation & $t'$ & 0.6 eV/{\AA}\\
    Fermi energy & $\varepsilon_F$ & 0.0 eV\\
    Atomic mass & $m$ & 197 a.m.u.\\
    Spring constant &$K$ & 2.0-8.0 eV/$\mathrm{\AA}^2$\\
    Temperature & $T$ & 4.2 $K$\\
    \hline
  \end{tabular}
  \caption{Model parameters used for metallic atomic wires.}
  \label{table1}
\end{IWCEtable}

The linear energy grid in principle has to cover the full
bandwidth (FBW) while at the same time it must have a resolution
fine enough to sample $\mathbf G^{r,\lessgtr}(\omega)$ and
$\mathbf\Sigma^{r,\lessgtr}_\alpha(\omega)$ well. For this model,
to resolve the fastest variations (caused by the Fermi function)
the grid point separation should be around $0.4$ meV or better at
a temperature of $T=4.2$ $K$. We find that calculations carried
out on an interval $[-\varepsilon_\mathrm{cut},
\varepsilon_\mathrm{cut}]$ converge quickly with
$\varepsilon_\mathrm{cut}$ to those of the FBW. As we show below
for a few representative cases, complete agreement is found when
$\varepsilon_\mathrm{cut}=0.1$ eV (which hence are used in the
calculations presented here). Over this narrow range we can
further apply the wide band limit (WBL) $\mathbf
\Sigma^r_\alpha(\omega)\approx \mathbf\Sigma^r_\alpha(\omega=0)$.
These simplifications reduce the computational load significantly.

\begin{IWCEfigure}
  \centering
  \includegraphics*[width=0.7\textwidth]{041016_L6_K2_NBvsFW.eps}
  \caption{Differential conductance and its derivative for a 6-atom wire
  with different values for the nearest neighbor spring constant $K$ in the externally
  damped limit ($n_\lambda\approx 0$). All 6 modes
  are included in this calculation.}
  \label{fig:damped}
\end{IWCEfigure}

The nonlinear conductance versus applied bias across a 6-atom wire
is shown (i) for the externally damped limit in
\Figref{fig:damped} and (ii) for the externally undamped limit in
\Figref{fig:undamped}. It is seen from \Figref{fig:damped} that
the conductance drop essentially happens at {\em one} particular
threshold energy. This energy is found to coincide with that of
the mode with highest vibrational energy, i.e.~the mode with
alternating bond length (ABL) character, which can also be
designated as the dominating one. This mode is further studied in
the externally undamped limit, \Figref{fig:undamped}, in which a
finite slope is observed beyond the threshold as well as a linear
increase in the mode population with bias (heating). Generally,
both figures show that the conductance drop {increases} while the
phonon threshold {decreases} when the spring constant is lowered.
This can be interpreted as an effect of straining the wire which
cause the bonds to weaken. Notice also the agreement in both
figures between the FBW and the WBL calculations, shown for the
case $K=2$ eV/${\mathrm\AA}^2$.

\begin{IWCEfigure}
  \centering
  \includegraphics*[width=0.7\textwidth]{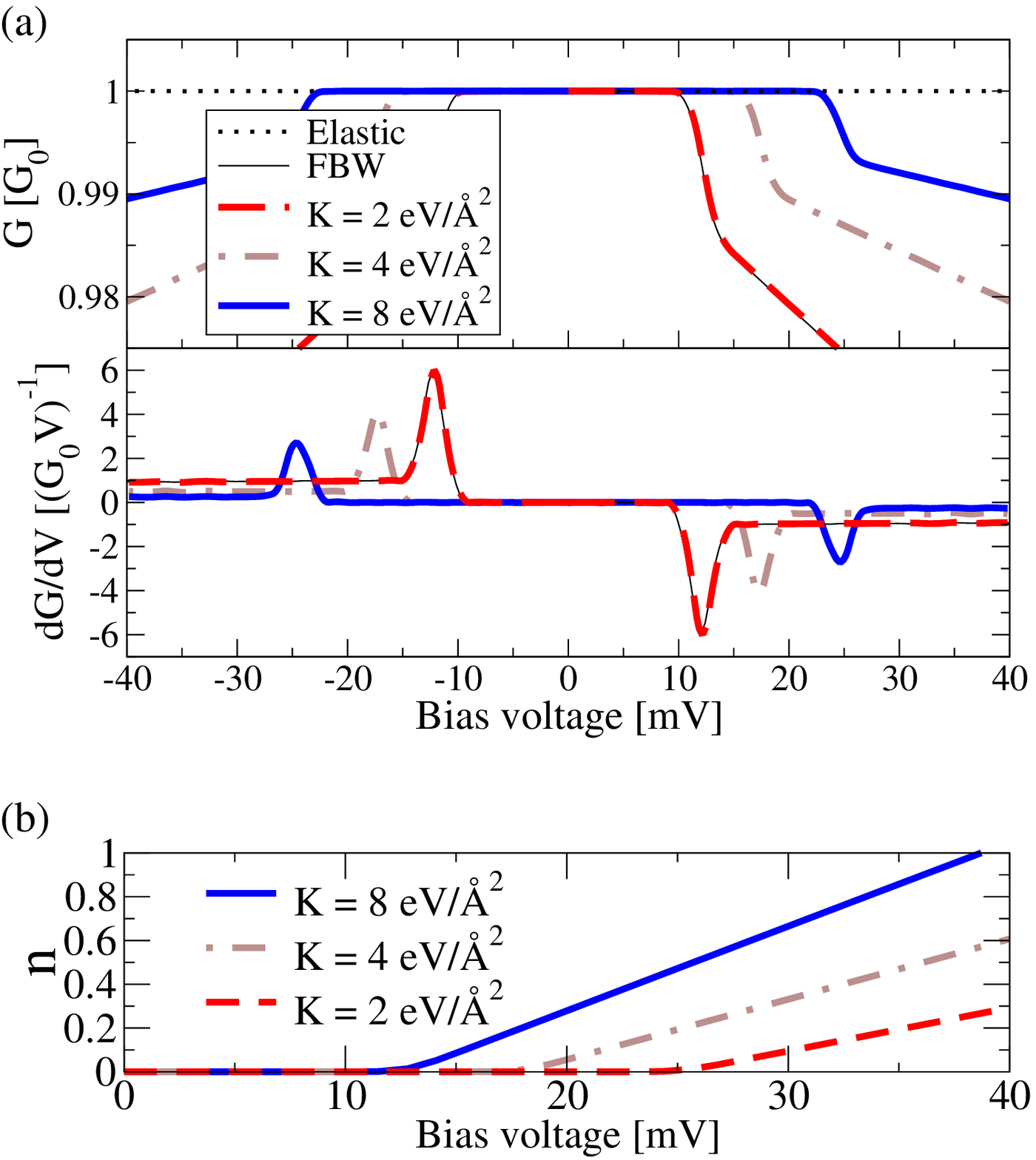}
  \caption{(\textsf{a}) Differential conductance and its derivative for a 6-atom wire
  with different values for the nearest neighbor spring constant $K$ in the externally
  undamped limit. Only the dominating mode is included in this calculation.
  (\textsf{b}) Mode occupation $n$ vs.~bias voltage.}
  \label{fig:undamped}
\end{IWCEfigure}

With our simple model we can easily handle longer wires. In
\Figref{fig:ScaleWithLength} we show a compilation of the
conductance drops and the conductance slopes for wires with length
up to $N=40$. The individual conductance plots all look
quantitatively much like those of \Figref{fig:damped} and
\ref{fig:undamped}. The important result is that these quantities
scale linearly with $N$. If one plots the conductance drop against
the inverse of mode energy (say, of the dominating mode) it is
found that the conductance drop also scales with $K$ as
$1/\Omega_\lambda$ (for fixed $N$), as one could speculate from
\Eqref{eq:couplings}.

\begin{IWCEfigure}
  \centering
  \includegraphics*[width=0.99\textwidth]{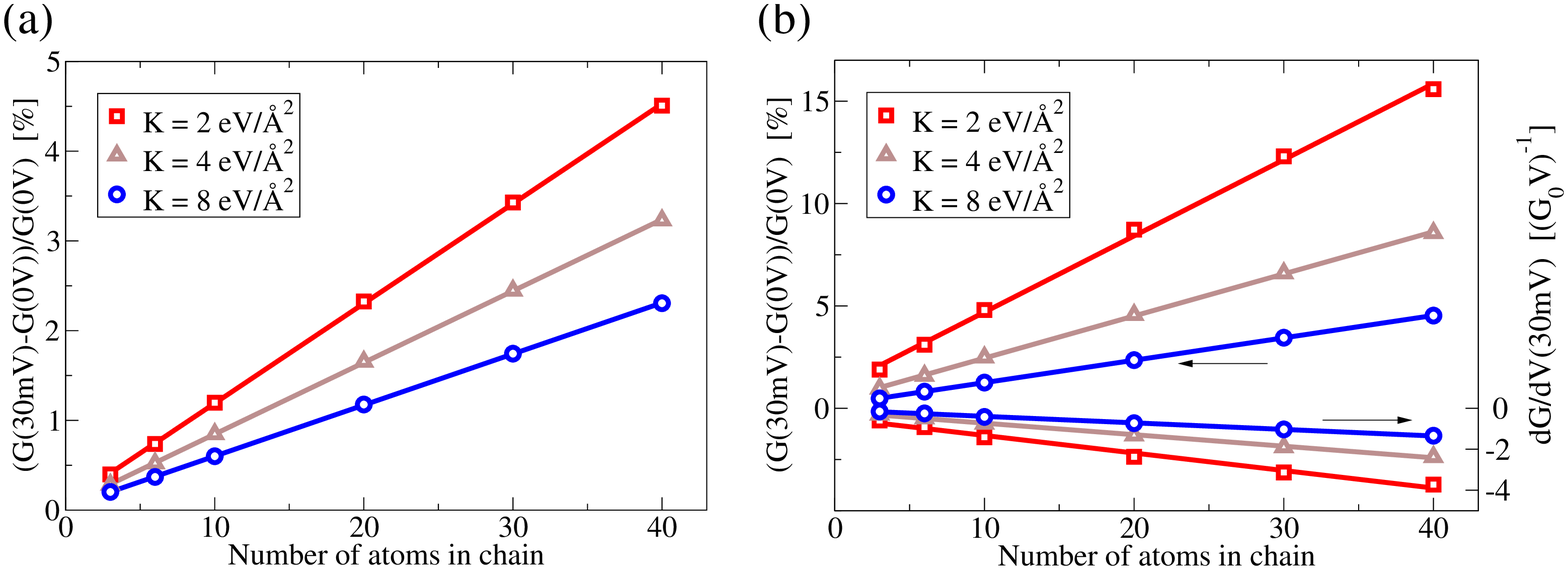}
  \caption{Compilation of the results obtained for different number of atoms
  in the wire (\textsf{a}) for the externally damped limit and (\textsf{b})
  for the externally undamped limit. The graphs show that the conductance drop and the conductance
  slope beyond threshold scale linearly with the length of the wire.}
  \label{fig:ScaleWithLength}
\end{IWCEfigure}

\section{Conclusions}
In conclusion, we have described a method to calculate inelastic
transport properties of an atomic-sized device connected between
metallic leads, based on NEGF techniques and SCBA for the
electron-vibration coupling. As a numerical example, we studied a
simple model for the transport through atomic Au wires. With a
single-orbital tight-binding description we illustrated the
significance of ABL mode character, and were able to explore even
very long wires. We further discussed the approximations related
to a representation on a finite energy grid.

As a final remark, and as we show elsewhere
\cite{FrederiksenEtAl.2004.ISLH}, the described method is also
well suited for a combination with full {\em ab initio}
calculations. The authors thank M.~Paulsson for many fruitful
discussions.

\end{multicols}
\end{document}